\documentclass[aps,pre,twocolumn,superscriptaddress,groupedaddress]{revtex4}  

\usepackage{graphicx}  
\usepackage{url}                  
\usepackage{dcolumn}       
\usepackage{bm}                  
\usepackage{amssymb}       
\usepackage{mathtools}     
\DeclarePairedDelimiter\ket{\lvert}{\rangle}
\usepackage{amsmath}    
\usepackage{pgf}                   
\usepackage{tikz,pgfplots}                   
\usetikzlibrary{shapes,arrows,calc}
\usetikzlibrary{arrows, decorations.markings}
\usepackage{lipsum}

\usepackage{soul}

\begin{document}

\title{Hybrid quantum-classical reservoir computing of thermal convection flow}
\author{Philipp Pfeffer}
\affiliation{Institut f\"ur Thermo- und Fluiddynamik, Technische Universit\"at Ilmenau, Postfach 100565, D-98684 Ilmenau, Germany}
\author{Florian Heyder}
\affiliation{Institut f\"ur Thermo- und Fluiddynamik, Technische Universit\"at Ilmenau, Postfach 100565, D-98684 Ilmenau, Germany}
\author{J\"org Schumacher}
\affiliation{Institut f\"ur Thermo- und Fluiddynamik, Technische Universit\"at Ilmenau, Postfach 100565, D-98684 Ilmenau, Germany}
\affiliation{Tandon School of Engineering, New York University, New York City, NY 11201, USA}
\date{\today}

\begin{abstract}
We simulate the nonlinear chaotic dynamics of Lorenz-type models for a classical two-dimensional thermal convection flow with 3 and 8 degrees of freedom by a hybrid quantum--classical reservoir computing model. The high-dimensional quantum reservoir dynamics are established by universal quantum gates that rotate and entangle the individual qubits of the tensor product quantum state. A comparison of the quantum reservoir computing model with its classical counterpart shows that the same prediction and reconstruction capabilities of classical reservoirs with thousands of perceptrons can be obtained by a few strongly entangled qubits. We demonstrate that the mean squared error between model output and ground truth in the test phase of the quantum reservoir computing algorithm increases when the reservoir is decomposed into separable subsets of qubits. Furthermore, the quantum reservoir computing model is implemented on a real noisy IBM quantum computer for up to 7 qubits. Our work thus opens the door to model the dynamics of classical complex systems in a high-dimensional phase space effectively with an algorithm that requires a small number of qubits.     
\end{abstract}

\maketitle

\section{Introduction}
Quantum computing (QC) and machine learning (ML) have changed our ways to process data fundamentally in the last years \cite{Jordan2015,LeCun2015,Preskill2018,Deutsch2020}. Quantum algorithms accelerated the data search \cite{Grover2001} or improved the sampling of probability distributions \cite{Arute2019,Wu2021}. These quantum advantages found already their way to various applications \cite{Orus2019,Ajagekar2019,Cao2019}, even though we are still in the era of noisy intermediate scale quantum (NISQ) devices that suffer from decoherence and are limited to qubit numbers $\sim 10^2$ with resulting shallow quantum circuit depths \cite{Altman2021}. 

Meanwhile, ML algorithms in the form of deep convolutional neural networks extract features effectively and classify big data bases \cite{Hinton2012,Ronneberger2015,Goodfellow2016,Fonda2019}. Quantum machine learning ports such methods to a quantum computer \cite{Biamonte2017,Ciliberto2017,Benedetti2019} with the prospect that particularly high-dimensional problems can be solved much faster than with their classical counterparts. This expectation arises from two facts, (i) the data space dimension grows exponentially as $2^n$ with the number of qubits $n$, the smallest unit of information in QC; (ii) the entanglement of qubits creates highly correlated tensor product states that can represent complex features in the data effectively \cite{Nielsen2010}. Thus{\color{red},} for example, quantum support vector machines are expected to have the potential to determine nonlinear decision boundaries of classification problems in high-dimensional quantum enhanced feature Hilbert spaces more efficiently \cite{Rebentrost2014,Schuld2019,Blank2020}. 

Recurrent neural networks (RNN) are specific ML algorithms with internal feedback loops which predict the time evolution of dynamical systems without knowing the underlying nonlinear ordinary or partial differential equations; they can be implemented either as gated RNNs in the form of long short-term memory networks \cite{Schmidhuber1997} or as reservoir computing models (RCM) \cite{Jaeger2001,Maass2002,Jaeger2004,Lukosevicius2009,Pathak2018}. As a consequence, RNNs have been used for the description of chaotic dynamics, fluid mechanical problems, and even turbulence \cite{Kutz2017,Srinivasan2019,Brunton2020,Pandey2020a}. RCMs were also applied to represent low-dimensional chaotic models, one-dimensional Kuramoto-Sivashinsky equations \cite{Lu2017,Vlachas2020}, or even turbulent Rayleigh-B\'{e}nard convection \cite{Pandey2020,Heyder2021,Pandey2022,Valori2022}. At the center of the RCM is the reservoir, a randomly initialized and fixed high-dimensionl network of perceptrons which is represented by an adjacency matrix. This specific implementation of an RNN requires only an optimization of the output layer, which maps the reservoir state back to the data space, and avoids costly backpropagation as required in most other ML algorithms \cite{Goodfellow2016}.                

In this work, we combine quantum algorithms with reservoir computing to a gate-based quantum reservoir computing model (QRCM) for a universal quantum computer to predict and reconstruct the dynamics of a thermal convection flow in the weakly nonlinear regime. The algorithm is of {\em hybrid quantum-classical nature} since the optimization of the output map is done by a classical ridge regression. The {\em quantum reservoir} is composed of a sequence of elementary single and two-qubit quantum gates which form a complex quantum circuit. Following the axioms of quantum mechanics, an elementary quantum gate performs a unitary transformation to a single- or two-qubit state. As a consequence, a highly entangled multi-quibit state will result. 

\begin{figure}
\includegraphics[scale=0.45]{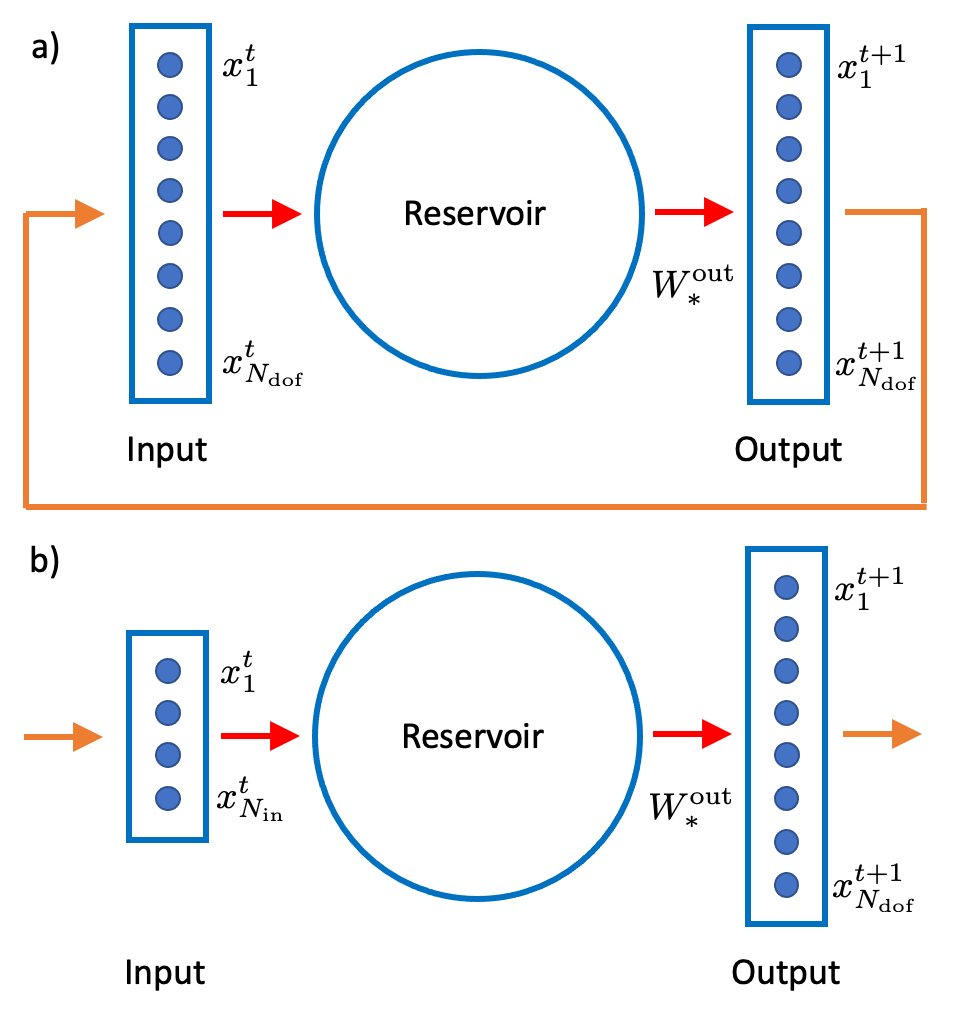}
\caption{Sketch of the two scenarios in which the reservoir computing model is run. (a) Closed-loop scenario for autonomous prediction of the dynamics. (b) Open-loop scenario for reconstruction of dynamics from continually available data. The matrix $W^{\rm out}_{\ast}$ stands for the classically optimized output layer. $N_{\rm dof}$ is the number of degrees of freedom of the dynamical system, $N_{\rm in}$ is the number of continually available components of the system state vector. The dimensionality of the reservoir is $N_{\rm res}\gg N_{\rm dof}$. For (a), $N_{\rm in} = N_{\rm dof}$ and for (b), $N_{\rm in}\ll  N_{\rm dof}$.} 
\label{fig0}
\end{figure}

Our first contribution is to demonstrate the feasibility of such a hybrid QRCM to describe the classical chaotic dynamics of a thermal convection flow on an actual NISQ device. The description of the thermal convection flow is based on Lorenz-type Galerkin models with $N_{\rm dof}\le 8$ degrees of freedom \cite{Lorenz1963,Howard1986,Thiffeault1996,Gluhovsky2002,Moon2017}. This class of models is directly derived from the Boussinesq equations of two-dimensional thermal convection between two impermeable parallel plates, heated uniformly from below and cooled from above with free-slip boundary conditions for the velocity field \cite{Koschmieder1993,Chilla2012}. Here, we explore QRCMs in two different modes of operation \cite{Lukosevicius2021}: 
\begin{enumerate}
\item {\em Closed--loop scenario}: a fully autonomous prediction of the temporal dynamics of all degrees of freedom of a Lorenz 63 model with $N_{\rm dof}=3$. This study is done with the ideal Qiskit quantum simulator \cite{Qiskit}, see the sketch in Fig. \ref{fig0}(a).  
\item {\em Open--loop scenario}: a reconstruction of the temporal dynamics of a Lorenz-type model with $N_{\rm dof}=8$. In this case, one or two degrees of freedom are continually fed into the quantum reservoir and the remaining degrees of freedom are obtained by the QRCM evolution, see Fig. \ref{fig0}(b). This investigation is done in two different ways. First, we reconstruct the whole model from a single degree of freedom ($N_{\rm in}=1$) by means of the open loop structure in an ideal Qiskit simulator. Secondly, we strongly reduce the number of quantum gates to even demonstrate the feasibility of QRCM on a real noisy IBM quantum computer ($N_{in}=2$). 
\end{enumerate}

Secondly, we directly compare the results of the QRCM to its classical counterpart for the same flow. We identify hyperparameters in both approaches that can be related to each other. Note that classical and quantum reservoir computing models differ essentially, which is primarily a consequence of the linearity and unitarity of the quantum dynamics \cite{Nielsen2010}.  

We demonstrate finally that a systematic reduction of the degree of entanglement in the quantum reservoir by a stepwise transition from a fully to a weakly entangled configuration reduces the performance of the present QRCM algorithm. More detailed, this is done by the division of an $n$-qubit reservoir state into blocks of entangled $p$-qubit states, so called $p$-blocks \cite{Josza2003}. The strong encoding capabilities of fully entangled quantum reservoirs are demonstrated in the present flow case by runs with qubit numbers $n<N_{\rm dof}$. Also, we show for the open-loop scenario, that the number of operations of the QRCM circuit can be scaled with $ {\cal O}(n)< {\cal O}(2^n)$ (where $2^n$ is the reservoir size).   

The research on quantum reservoir computing models proceeds along two major frameworks \cite{Markovic2020,Mujal2021}. \\
(1) The dynamics of an interacting boson or fermion many-particle quantum system is investigated in the {\em analog} framework, which is characterized by a Hamiltonian subject to a unitary time evolution. These systems have been established in the form of spin ensembles \cite{Fujii2017,Nakajima2019,Kutvonen2020,Sakurai2021}, circuit quantum electrodynamics \cite{Khan2021}, arrays of Rydberg atoms \cite{Araiza2021}, or networks of linear quantum optical oscillators \cite{Nokkala2021}. In refs. \cite{Martinez2021,Xia2022}, the phase transition from a thermalized to a localized many-particle quantum reservoir was studied in respect to the echo state property. The latter describes the ability of the (quantum) reservoir to forget its initial conditions. It is shown that thermalized quantum reservoirs close to the phase transition boundary, for which all spins or oscillators are still strongly entangled, show the best performance for nonlinear learning tasks. A further way to establish a quantum reservoir is by a single nonlinear oscillators \cite{Govia2021}. 

(2) The {\em digital gate-based} framework, which sets the stage for the present work, uses circuits composed of universal quantum gates to build a quantum reservoir on NISQ devices \cite{Chen2020,Dasgupta2020,Suzuki2022}. These configurations have been applied for the one-step prediction of nonlinear auto-regressive moving-average (NARMA) time series or solutions of the nonlinear Mackey-Glass time-delay differential equation. Here we extend the applications to classical nonlinear dynamical systems with up to 8 degrees of freedom. Furthermore, we apply a reservoir update that blends a linear and a nonlinear activation term, as frequently done in classical reservoir models. 

Our work opens the door for the application of quantum machine learning as a reduced-order dynamical model of a higher-dimensional classical complex dynamical nonlinear system. The study thus adds a further proof-of-concept for the potential use of quantum algorithms in studying turbulent flows. 

The outline is as follows. In section II, we present the thermal convection flow model; technical details are collected in appendix A. Section III is dedicated to the closed loop scenario. In section IV the complexity of the quantum machine learning task is enhanced to the 8-dimensional model for which we apply an open--loop QRCM. We summarize our work and give a brief outlook in section V.  Appendices B, C, and D provide additional material on $n$-qubit quantum states, the classical reservoir computing framework, and benchmarks of the QRCM with different leaking rates.

\begin{figure*}
\includegraphics[scale=0.52]{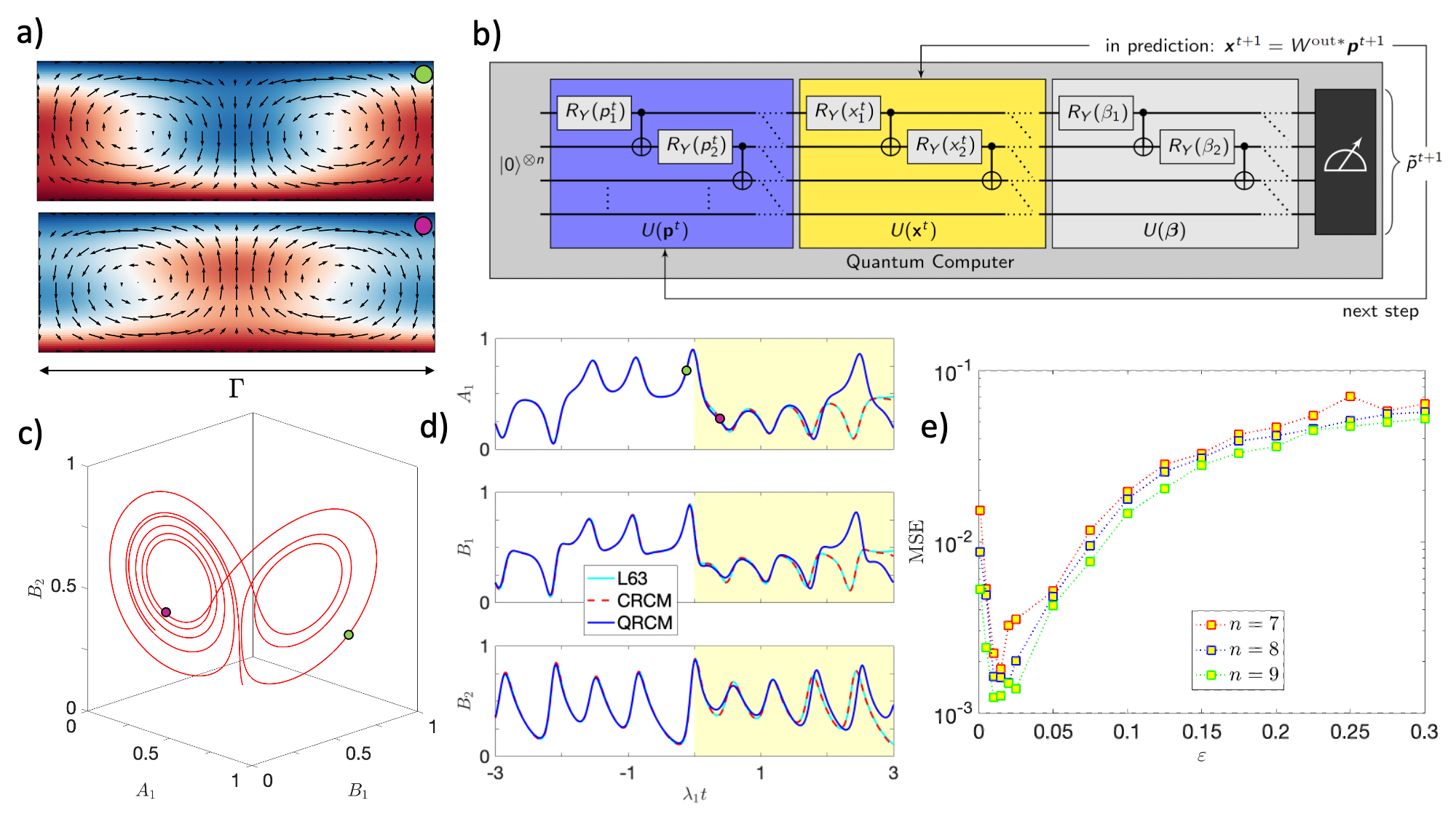}
\caption{Quantum reservoir computing model (QRCM) for the Lorenz 63 dynamical system. (a) Two instantaneous convection flow states which display the velocity vector field $(u_x,u_z)$ together with the total temperature field $T$ as a colored background (blue for $T_{\rm top}$ and red for $T_{\rm bot}$). (b) Circuit diagram for the QRCM. Here $(x_1,x_2,x_3)=(A_1,B_1,B_2)$. The three groups of unitary operations are indicated by differently colored boxes. The scaling of input parameters of the $R_Y$-rotation gates is specified in the text. (c) Trajectory plot of the Lorenz 63 system in the phase space which is spanned by one stream function mode $A_1(t)$ and two temperature modes, $B_1(t)$ and $B_2(t)$, i.e., $N=1$ and $M=2$. (d) Comparison of classical and quantum reservoir computing in the prediction phase (yellow background). Time is rescaled by the largest Lyapunov exponent $\lambda_1=0.9056$, see e.g. ref. \cite{Geurts2020}. The two flow configurations in (a) are also indicated in panels (c,d). (e) Mean squared error (MSE) as a function of the leaking rate $\varepsilon$ and the number of qubits $n$. The model parameter are $\sigma=10$, $b=8/3$, and $r=28$. All displayed QRCM runs are done with the Qiskit simulator.}
\label{fig1}
\end{figure*}

\section{Thermal convection flow} 
We start with a compact description of the flow. The thermal convection flow consists of a two-dimensional fluid layer which is heated uniformly from below with a temperature $T_{\rm bot}$ and cooled from above with $T_{\rm top}$ thus giving $\Delta T=T_{\rm bot}-T_{\rm top}>0$.  The convection flow domain is $A=[0,\Gamma]\times [0,1]$. The velocity ${\bm u}({\bm x},t)=(u_x(x,z,t), u_z(x,z,t))$ and (total) temperature $T(x,z,t)$ are coupled by the balances of mass, momentum, and energy. The fluid is incompressible and the mass density $\rho$ depends linearly on $\theta$ in the buoyancy term only. This is known as the Boussinesq approximation in thermal convection \cite{Chilla2012}. The total temperature is decomposed into $T(x,z,t)=1-z+\theta(x,z,t)$ where $T_{\rm eq}(z)=1-z$ is the static equilibrium profile and $\theta(x,z,t)$ is the temperature deviation. The non-dimensional equations are then given by
\begin{align}
{\bm \nabla} \cdot {\bm u}&= 0 \,,
\label{eq:NS1}\\
\frac{\partial {\bm u}}{\partial t} + ({\bm u} \cdot {\bm \nabla}) {\bm u} &= - {\bm \nabla} p + \sigma {\bm \nabla}^2 {\bm u} + {\rm Ra}\, \sigma\, \theta {\bm e}_z \,,\label{eq:NS2b}\\
\frac{\partial \theta}{\partial t} + ({\bm u}\cdot {\bm \nabla})\theta &= {\bm \nabla}^2 \theta + u_z\,
\label{eq:NS3}
\end{align}
where we use the kinematic pressure $p$. The Rayleigh number ${\rm Ra}$ and the Prandtl number $\sigma$ are the two parameters that characterize the strength of the thermal driving via the temperature difference $\Delta T$ and the ratio of momentum to temperature diffusion, respectively.  The boundary conditions in $x$-direction are periodic. At $z=0,1$, one takes
\begin{align}
u_z\big|_{z=0,1}=0,\quad \frac{\partial u_x}{\partial z}\Bigg|_{z=0,1}=0\quad\mbox{and}\quad \theta\big|_{z=0,1}=0\,.\label{bc1}
\end{align}
They correspond to isothermal, impermeable, free-slip walls. Incompressibility and two-dimensionality allow to reduce the velocity vector field further to a scalar stream function $\zeta(x,z,t)$ by 
\begin{equation}
u_x=-\frac{\partial \zeta}{\partial z} \quad \mbox{and} \quad u_z=\frac{\partial \zeta}{\partial x}\,. 
\end{equation}
This ansatz satisfies \eqref{eq:NS1} automatically and the equations of motion \eqref{eq:NS2b}--\eqref{eq:NS3} are now given by 
\begin{align}
\frac{\partial \nabla^2\zeta}{\partial t}&=\frac{\partial \zeta}{\partial z}  \frac{\partial \nabla^2\zeta}{\partial x}-\frac{\partial \zeta}{\partial x}  \frac{\partial \nabla^2\zeta}{\partial z} + \sigma \nabla^4\zeta +{\rm Ra} \sigma\frac{\partial \theta}{\partial x}\,,\label{2d1}\\   
\frac{\partial \theta}{\partial t}&=\frac{\partial \zeta}{\partial z}  \frac{\partial \theta}{\partial x}-\frac{\partial \zeta}{\partial x}  \frac{\partial \theta}{\partial z} + \nabla^2\theta +\frac{\partial \zeta}{\partial x}\label{2d2}\,,
\end{align}
with boundary conditions in the vertical direction
\begin{align}
\zeta\big|_{z=0,1}=0,\quad \frac{\partial^2\zeta}{\partial z^2}\Bigg|_{z=0,1}=0\quad\mbox{and}\quad \theta\big|_{z=0,1}=0\,.\label{bc}
\end{align}
Equations \eqref{2d1} and \eqref{2d2} are then reduced by an expansion into trigonometric Fourier modes which satisfy the boundary conditions for the stream function and temperature and encode the spatial structure of the thermal convection flow, see appendix A for further technical details. A subsequent truncation to $N$ and $M$ real time-dependent amplitudes is done for the stream function, $\{A_1(\tau), \dots, A_N(\tau)\}$, and the temperature, $\{B_1(\tau), \dots, B_M(\tau)\}$, respectively. 

This step leads to a class of low-dimensional Lorenz-type Galerkin models of the thermal convection flow starting with the original three-dimensional Lorenz 63 model \cite{Lorenz1963} for $N=1$ and $M=2$ (where $N_{\rm dof}=N+M$). The resulting coupled nonlinear system of ordinary differential equations is given by  
\begin{align}
\frac{dA_i}{d\tau}&=F_i(A_j, B_k, \sigma, r, b)\,,\label{Lo1}\\
\frac{dB_k}{d\tau}&=G_k(B_l, A_i, \sigma, r, b)\,, \label{Lo2}
\end{align}
for $i,j=1\dots N$ and $k,l=1\dots M$. Here, $\sigma$ is again the Prandtl number, $r$ the {\em relative} Rayleigh number, and $b$ an aspect ratio parameter, see appendix A. Furthermore, $F_i$ and $G_k$ are quadratic nonlinear functions of the amplitudes $A_i(\tau)$ and $B_k(\tau)$. We will consider two implementations, the Lorenz 63 model (L63) \cite{Lorenz1963} with $N=1$ and $M=2$ and an extended 8-dimensional model \cite{Gluhovsky2002} with $N=M=4$ that introduces shear in the flow and causes tilted convection rolls and shearing motion. It thus displays a more complex fluid motion further away from the primary instability point at $r=1$ or $Ra_c=27\pi^4/4$ \cite{Koschmieder1993}. 

Figures \ref{fig1}(a) shows two instances of the temperature and velocity fields with the counter-rotating circulation rolls that cause a rise of warm and a descent of cold fluid. These two flow states corresponds to trajectory points of L63 in each of the two butterfly-like wings in Fig. \ref{fig1}(c).    

\section{Closed-loop scenario for three-dimensional Lorenz model} 
\subsection{Quantum and classical reservoirs}
The design of our time-discrete and gate-based QRCM builds on a $n$-qubit tensor product state at time $t$. In appendix B, we provide a compact primer on qubits, tensor product spaces, and entangled or separable states. The $n$-qubit state in Dirac notation \cite{Nielsen2010} is given by
\begin{equation}
    |{\bm \psi}^t\rangle=\sum_{k=1}^{N_{\rm res}} a^t_k |k\rangle \quad\mbox{with}\quad a_k^t\in \mathbb{C}\,,
    \label{psi}
\end{equation}
with $N_{\rm res}=2^n$. Here $|k\rangle$ is the standard basis of the $n$-qubit quantum register. The measured probabilities $p_k$ are given by
\begin{equation}
p_k^t = {|a_k^t|}^2
\label{Rho_and_P}
\end{equation} 
with $a_k^t$ from eq. \eqref{psi}. Here, we have $2^n$ probabilities; $2^n-1$ of them are linear independent since they have to sum up to 1. The reservoir state evolves from time $t$ to $t+\Delta t$ with a fixed time step width $\Delta t$ as follows. First, the dynamical part is updated by three blocks of unitary linear transformations
\begin{equation}
|\tilde{\bm \psi}^{t+1}\rangle= U({\bm \beta})U(4\pi{\bm x}^t)U(4\pi{\bm p}^t) \ket{0}^{\otimes n} , \label{unitary}
\end{equation}
with random rotation angles ${\bm \beta}=(\beta_1, ..., \beta_n)$, reservoir state probability amplitudes ${\bm p}^t=(p_1^t, ..., p_{2^n}^t)$, and the past system state vector ${\bm x}^t=(x_1^t, ..., x_{N+M}^t)$, the latter of which summarizes $(A_1,...,B_M)$. We simplify the notation in \eqref{unitary} by switching from $t+\Delta t$ to $t+1$ (or later $t+m$ with $m\in\mathbb{N}$). The initial $n$-qubit state vector $\ket{0}^{\otimes n}$ implies that every qubit is in the basis state $\ket{0}$. With eq. (\ref{Rho_and_P}) for the probability amplitudes $\tilde{p}_k^{\ t+1}$ which are obtained from $|\tilde{\bm \psi}^{t+1}\rangle$, the RCM update step outside the quantum reservoir is given by the following iteration 
\begin{equation}
p_k^{t+1} = (1-\varepsilon) \ p_k^{t} + \varepsilon \ \tilde{p}_k^{\ t+1}\,.
\label{MAIN_Eq}
\end{equation}
The update rule thus contains two terms, a first linear memory term and a second nonlinear activation term. The nonlinearity is connected to the classical data loading as will be discussed in the next subsection. Equation \eqref{MAIN_Eq} contains a leaking rate $0\le \varepsilon\le 1$ that blends both terms. In the classical reservoir computing model, the update of the reservoir state ${\bm \psi}_c^{t}$ would be given by  
\begin{equation}
	\label{rcm_train0}
	{\bm \psi}_c^{t+1}= (1-\varepsilon){\bm \psi}_c^t + \varepsilon\tanh\left[W^{\rm in}{\bm x}^t + W^{\rm r} {\bm \psi}_c^{t}\right]\,
\end{equation}
which reveals a similar structure to \eqref{MAIN_Eq}. Here, $W^r$ is the reservoir matrix and $W^{\rm in}$ the input matrix. More details are provided in appendix C. A leaking rate of $\varepsilon=1$ implies that only a nonlinear activation by the hyperbolic tangent is present -- a mode in which several, but not all classical reservoir computing models are operated \cite{Jaeger2001,Jaeger2002}. Inubushi and Yoshimura \cite{Inubushi2017} term this split the memory-nonlinearity trade-off since the nonlinear activation term typically will degrade the memory of the system. 

Differently to the analog framework of quantum reservoir computing \cite{Fujii2017,Nakajima2019,Kutvonen2020,Sakurai2021,Khan2021,Araiza2021,Nokkala2021,Martinez2021,Xia2022} that processes the state without external memory in the reservoir, we add external memory by the first term in \eqref{MAIN_Eq}. An improved performance of the present hybrid quantum-classical reservoir computing model for $\varepsilon <1$ in comparison to one with $\varepsilon=1$ is demonstrated in appendix D for two common benchmark cases.

\subsection{Classical data loading and reservoir state evolution}
We now specify the loading procedure of the classical data into the quantum reservoir. The unitary transformations of eq. \eqref{unitary} consist of single-qubit rotation gates $R_Y$ and subsequent two-qubit controlled NOT (in short CNOT) gates. The $R_Y$-gate is defined by $$ R_Y(x) = \begin{pmatrix} \cos(x/2) & -\sin(x/2) \\ \sin(x/2) & \;\;\;\cos(x/2) \end{pmatrix}.$$

Figure \ref{fig1}(b) shows the corresponding circuit diagram of the quantum reservoir which consists of three circuit blocks as lined out in eq. \eqref{unitary}. The first block of unitary transformations $U(4\pi{\bm p}^t)$ loads the reservoir state probability amplitudes of the previous time step $t$ (indicated as the blue box). This is done by rotation gates $R_Y(4\pi p^t_k)$. In the circuit diagram of Fig. \ref{fig1}, these $2^9=512$ probabilities with $0\le p_k^t\le 1$ are summarized to a vector to keep the notation less crowded. For this as for all the following blocks, the combination of $R_Y$ and CNOT gates is continued until the last qubit is reached. There, the CNOT is applied to the previous qubit and if not yet finished, the constructor starts at the upper qubit again.

The application of an $R_Y$ rotation gate, which is parametrized by the input value, is a {\em nonlinear} operation in terms of the amplitudes \cite{Dasgupta2020,Govia2022}. It can be considered as an analogy to the nonlinear activation in a classical RCM, e.g., by $\tanh(\cdot)$, see also table \ref{table1} where we summarize hyperparameters of the classical and quantum RCMs. Note also that all $2^9$ amplitudes $p_k^t$ are loaded into the reservoir in this case.

Similarly, the degrees of freedom $x_i^t$ at time $t$ are loaded into the quantum reservoir in the second block $U({4\pi\bm x}^t)$ before the model advances further (indicated as the yellow box) to the last block (indicated as the gray box). This third block $U({\bm \beta})$ performs additional rotations by angles $\beta_i$. It stands for a unitary evolution step of the full reservoir state which enhances the entanglement and randomization. The rotation angles $\beta_i$ are sampled initially in a reproducible way from a uniform distribution between 0 and $2\pi$, which corresponds to a random seed initialization of a classical reservoir.

The loading of the full reservoir state, which becomes exponentially more costly, was necessary to obtain the reported prediction horizons for the closed-loop scenario. This is discussed in the next subsection.
\begin{table*}
    \renewcommand{\arraystretch}{1.1}
    \centering
    \begin{tabular}{lccccc}
        \hline\hline
        Quantity & $\;\;$ Classical RCM $\;\;$& Optimal value & $\quad$ & $\;\;$ Quantum RCM $\;\;$ & Optimal value \\
        \hline
        Reservoir dimension & $N_{\rm res}$ & 512 & & $N_{\rm res}=2^n$ & 512\\
        Leaking rate & $\varepsilon$ & 0.12 & & $\varepsilon$ & 0.05 \\
        Spectral radius of reservoir & $\rho(W_r)$ & 1.01 & & $\rho(U)$ & 1.0 \\
        Reservoir state at time $t$& ${\bm \psi}^t\in \mathbb{R}^{N_{\rm res}}$& & & $|{\bm \psi}^t\rangle\in \mathbb{C}^{N_{\rm res}}$ & \\
        Training steps & $N_{\rm train}$ & 2000 & & $N_{\rm train}$ & 2000 \\
        Reservoir model nonlinearity & $\tanh(\cdot)$ & & & $R_Y(\cdot)$ & \\
        \hline\hline
    \end{tabular}
    \caption{Comparison of classical and quantum reservoir computing models. Different essential quantities including optimal hyperparameters for the Lorenz 63 model in the closed-loop scenario are listed. The spectral radius $\rho(W^r)$ in the quantum case is always equal to 1 since unitary transformations are norm-preserving. The number of qubits is $n$. Two additional hyperparameters are used in the classical RCM: a reservoir density $D=0.2$ which determines the percentage of active nodes in the matrix $W^r$ and an additional Tikhonov regularization term with a parameter $\gamma=10^{-1}$ in the cost function $C(W^{\rm out})$, see appendix C.}
    \label{table1}
\end{table*}

\subsection{Quantum reservoir readout and classical optimization}
A projection-valued measure in the standard basis of the Pauli-$Z$ operator provides the probabilities $p_k^t$ from $K\gg 2^n$ independent circuit simulations, known as shots. These probabilities are mapped to the updated dynamical system state by the output matrix,    
\begin{align}
x_i^{t} &= \sum_{k=1}^{2^n} W^{{\rm out}\ast}_{ik} p^t_k\,,  
\label{outp}
\end{align} 
with the optimized weights which are summarized in the matrix $W^{{\rm out}\ast}\in \mathbb{R}^{(N+M)\times N_{\rm res}}$. We note once more that the output matrix is optimized by a classical algorithm similar to the classical RCM case. This optimization seeks a minimum of the cost function $C(W^{\rm out})$ which is given in appendix C. 

Panel (d) of Fig. \ref{fig1} and table \ref{table1} compare the classical and quantum RCM with the numerical simulation of the equations of motion obtained by a 4th-order Runge-Kutta method. The integration time is rescaled by the largest Lyapunov exponent of the system, $\lambda_1=0.9056$, which quantifies the deterministic chaos of the model \cite{Geurts2020}. The training phase comprises $N_{\rm train}=2000$ time steps, both for the classical and quantum case. The first 50 time steps out of $N_{\rm train}$ are used for the washout of the initial reservoir state. For times $t\ge 0$ the reservoirs are exposed to unseen test data predicting the dynamics autonomously. It is seen that the prediction horizon of the QRCM with 9 qubits is about $1.5 \lambda_1 t$ in this example. This result remains nearly the same for different reservoir seeds. Either the approximation of the L63 model by reservoir dynamics or the additional white noise in the Qiskit simulator will cause a switch of the trajectory into the other wing of the butterfly-like Lorenz attractor. Note that the classical RCM (CRCM) prediction will also deviate from the ground truth at about $3\lambda_1 t$ which is not shown here. The leaking rates in this example are given in table \ref{table1}.

Two hyperparameters are varied, the leaking rate $\varepsilon$ and the number of qubits $n$ that determines the reservoir dimension $N_{\rm res}$.  We identify a minimum of the cost function in the form of a mean squared error (MSE) around $\varepsilon=0.025$. This is the statistical minimum while the single best representation shown in Fig. \ref{fig1}(d) has $\varepsilon=0.05$, as stated in table \ref{table1}. The larger the number of qubits the smaller MSE, although the improvements in the Qiskit simulations remain small (and thus the difference of the displayed to the optimal case). Note also that each data point for the MSE in Fig. \ref{fig1}(e) is obtained as an average over 50 different random seeds of the quantum reservoir, i.e., 50 different random vectors with angles $\beta_i$. A small leaking rate implies that the reservoir dynamics is memory-dominated blended with a small nonlinear contribution \cite{Inubushi2017}, as we detailed already in Sec. III A. 

The Tikhonov regularization parameter $\gamma$ which is added to the cost function to avoid overfitting was set to $\gamma=0$ in the present as well as in the NISQ device runs. The noise in the ideal quantum simulator and the decoherence in the NISQ device have a regularizing effect, see discussions of this aspect in ref. \cite{Jaeger2002} for classical and in \cite{Suzuki2022} for quantum devices. We will come back to this hyperparameter in the next section.  All details on the classical reservoir computing model, the hyperparameters, and the cost function are in appendix C in order to keep the work self-contained.   

The quantum reservoir readout requires $K$ projective measurements of the $n$-qubit reservoir state on identically prepared quantum systems which might be costly in comparison to the CRCM case. However, a readout of the data has to be done in the classical case as well. Throughout this study, we used $K=2^{10+n}$ shots to reduce the measurement noise. A further increase of the number of shots did not reduce the MSE. Since the NISQ devices allow a maximum number of $K=8192$ shots only, batch jobs with a pre-defined Qiskit function (combine\underline{\hspace{0.3cm}}counts) were used. 

In the closed-loop scenario, we conduct these measurements after each step to monitor the time evolution at equidistant instants from $t$ to $t_m=t+m$ ($m>1$, $m\in \mathbb{N}$). This is the same procedure as in the classical algorithm such that both can be compared to each other. With an increasing number of qubits the state vector grows exponentially. Larger qubit numbers can also require a number of shots $K$ for the quantum simulator or quantum device that goes beyond our presently suggested one. In the open-loop scenario, which will be discussed in the next section IV, we advance the system from $t$ to $t+1$ only to obtain the results. This step always closes with a readout in the form of a measurement.
\begin{figure*}
\includegraphics[scale=0.6]{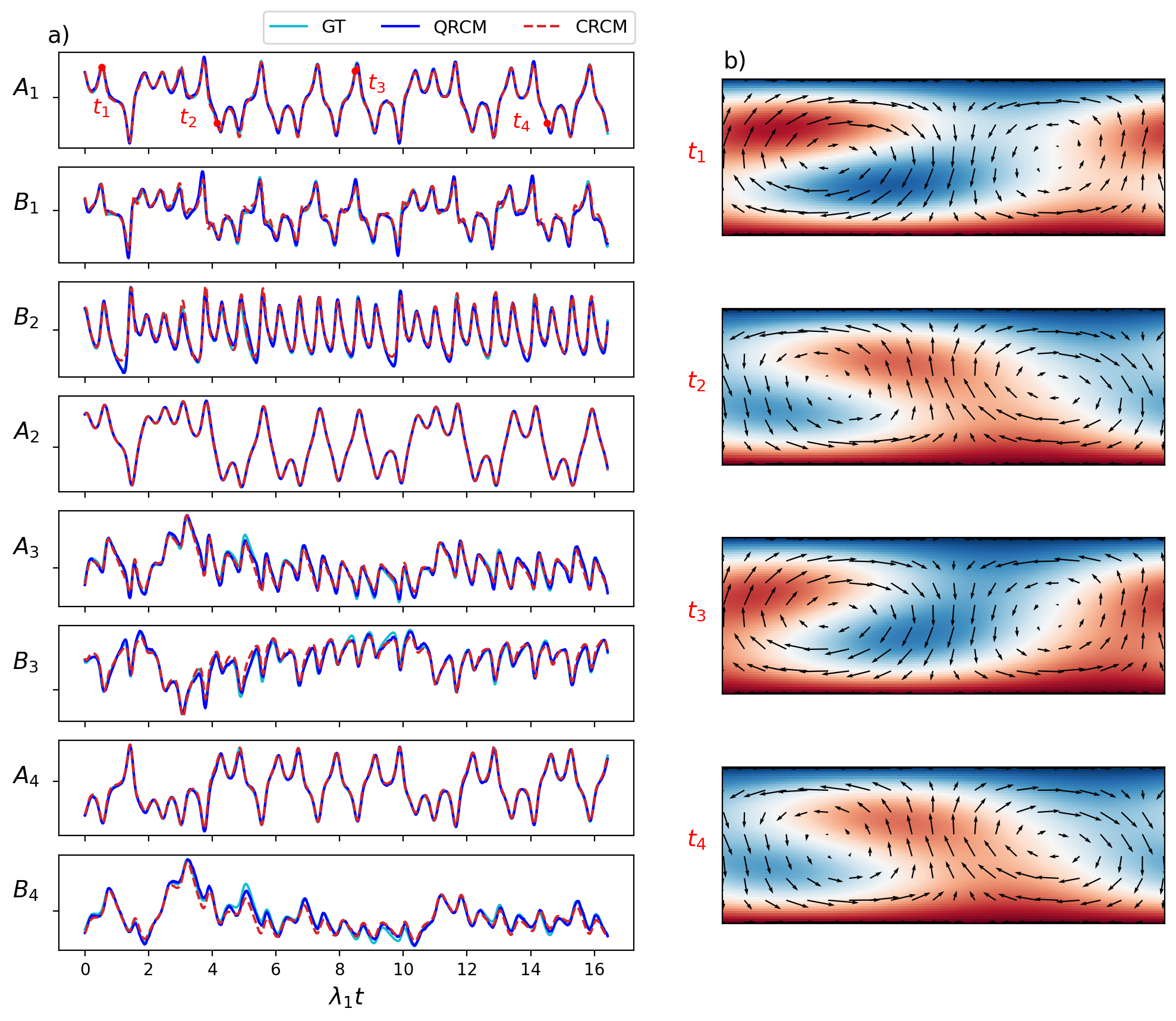}
\caption{Comparison of the 8-dimensional Lorenz-type model, time-integrated system of equations as ground truth (GT), classical reservoir computing model (CRCM), and quantum reservoir computing model (QRCM). Panel (a) displays the time evolution of all variables. (b) Reconstruction of the flow and temperature fields at times $t_1$ to $t_4$ which are indicated in $A_1(t)$. The model parameters are $\sigma=10$, $b=8/3$, and $r=28$. Here, $N=4$ and $M=4$. Mode $A_4$ is the only input and always given accurately into each reservoir.}
\label{fig2}
\end{figure*}

\section{Open-loop scenario for 8-dimensional Lorenz-type model}
\subsection{Quantum reservoir with one continually available degree of freedom}
We proceed from the standard L63 model to an extension with  8 degrees of freedom, which is listed and explained further in appendix A. As shown by Gluhovsky {\it et al.} \cite{Gluhovsky2002}, this extension can be decomposed into subgroups, so-called gyrostats. The model conserves total energy and vorticity. They are given by 
\begin{equation}
    E(t)=\frac{1}{2A}\int_A ((\nabla\zeta)^2-z\theta) dA\,,
\end{equation}
with a kinetic and potential energy term and
\begin{equation}
\Omega(t)=\frac{1}{A}\int_A \omega dA\,,   
\end{equation}
with the vorticity $\omega=-\nabla^2\zeta$ and the convection domain size $A$. The open-loop scenario of the QRCM implies that a subset of the degrees of freedom will remain continually available in the reconstruction phase after the training phase. In this subsection, we will take $N_{\rm in}=1$ which will be $A_4(t)$. The leading Lyapunov exponent was computed numerically by a method proposed in~\cite{Sprott2003} and turned out to be $\lambda_1= 0.825$.
  
Figure \ref{fig2} displays the results for the 8--dimensional Lorenz-type model  which receives the time series $A_4(t)$ to reconstruct the remaining 7 degrees of freedom of the thermal convection flow model. Panel (a) compares the times series of the ground truth (GT) with the results of a CRCM, and a QRCM which was run on $n=7$ qubits on an ideal Qiskit simulator. We see that the data remain closely together for the displayed interval of more than 16 Lyapunov time units. The QRCM runs again through a training phase of $N_{\rm train}=2000$ integration time steps with a leaking rate $\varepsilon=0.05$ after an initial washout of 50 time steps. Figure \ref{fig2}(b) displays the reconstructed convection flow at four instants. The 8-dimensional model incorporates the shearing modes which are missing in the lower-dimensional Lorenz 63 model and lead to tilted convection cells, as can be seen in the panels.

\begin{figure}
    \centering
    \includegraphics[scale=0.45]{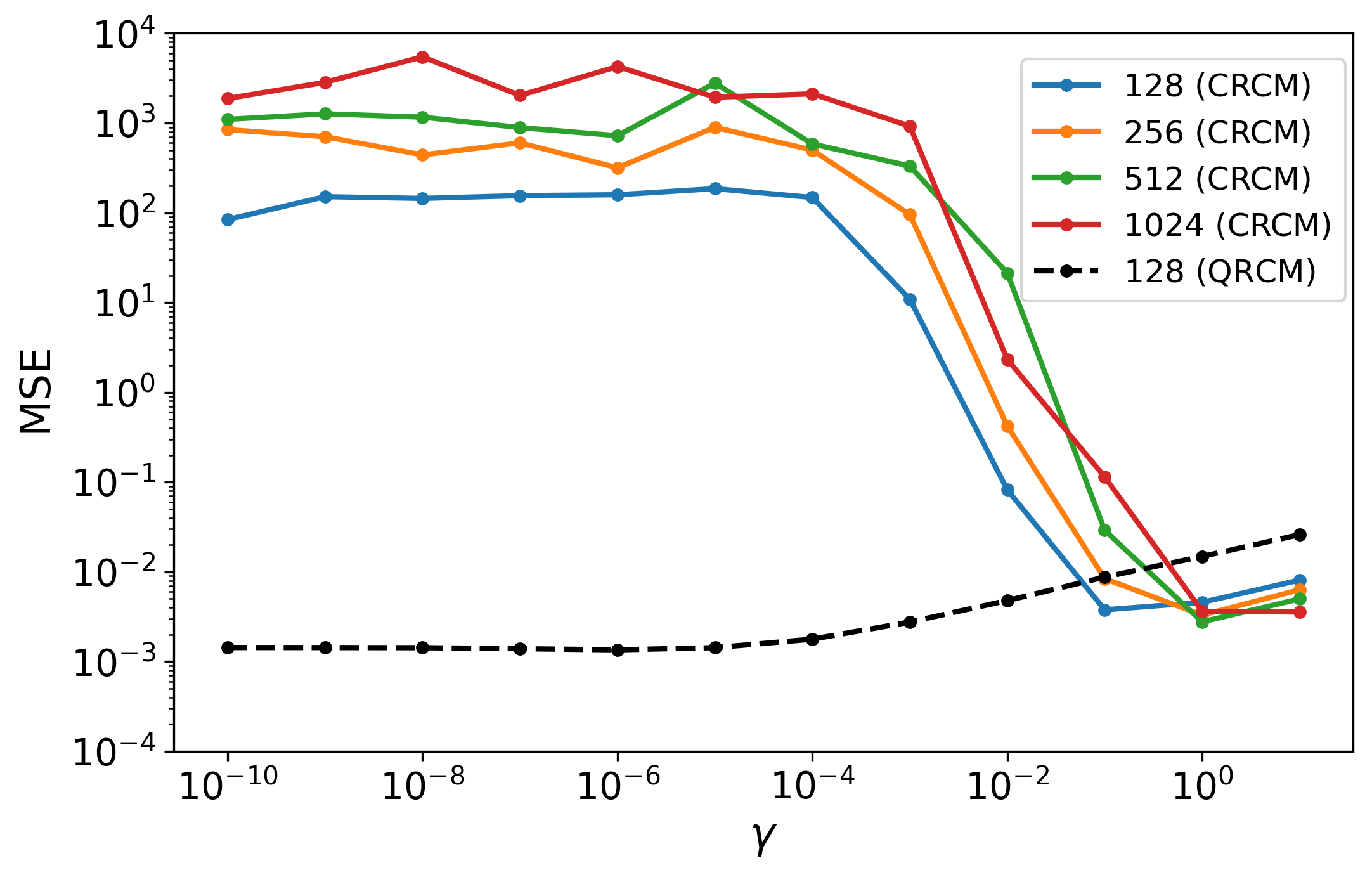}
    \caption{Comparison of classical and quantum reservoir computing models in the reconstruction phase for different regularization parameters $\gamma$ which have been increased by factors of 10 from $10^{-10}$ to 10. Each data point corresponds to the median of 30 different random reservoir realizations. The remaining hyperparameters remain fixed throughout this study. }
    \label{fig_reg}
\end{figure}

The dimension of the quantum reservoir is $N_{\rm res}=128$, while the one of the CRCM in Fig. \ref{fig2} is $N_{\rm res}=512$. We now compare the mean squared error (MSE) as a function of the dimension of the CRCM $N_{\rm res}$ and the regularization parameter $\gamma$ in Fig. \ref{fig_reg}. We also show the behavior of the QRCM $N_{\rm res} = 128$ for comparison. The MSE is given as
\begin{equation}
    \mbox{MSE}=\frac{1}{T_{\rm test}}\sum_{t=1}^{T_{\rm test}}|{\bm x}^t-{\bm x}_{\rm tg}^t|^2\,,
\end{equation}
see also eq. \eqref{outp} and $T_{\rm test}=2000$. Subscript tg stands for target and denotes the ground truth (GT) which is obtained by time integration of the model eqns. \eqref{Lo1} and \eqref{Lo2}. The MSE of the CRCM is large for all reservoir dimensions when $\gamma$ is very small, but improves as $\gamma$ rises. The minimal MSE for the CRCM is $\approx 2.8\times 10^{-3}$ at $\gamma=1$ and $N_{\rm res}=512$ while the minimal value of the QRCM is $\approx 1.36\times 10^{-3}$. The MSE of the QRCM remains practically unchanged for $\gamma\lesssim 10^{-3}$ and starts to grow then moderately. As we discussed in the last section already, the noise of the quantum algorithms seems to provide already enough regularization. We can conclude that the MSE of the QRCM is found to be fairly close to that of the CRCM with a reservoir dimension that is reduced by a factor of 4. 

\begin{figure*}
\includegraphics[scale=0.5]{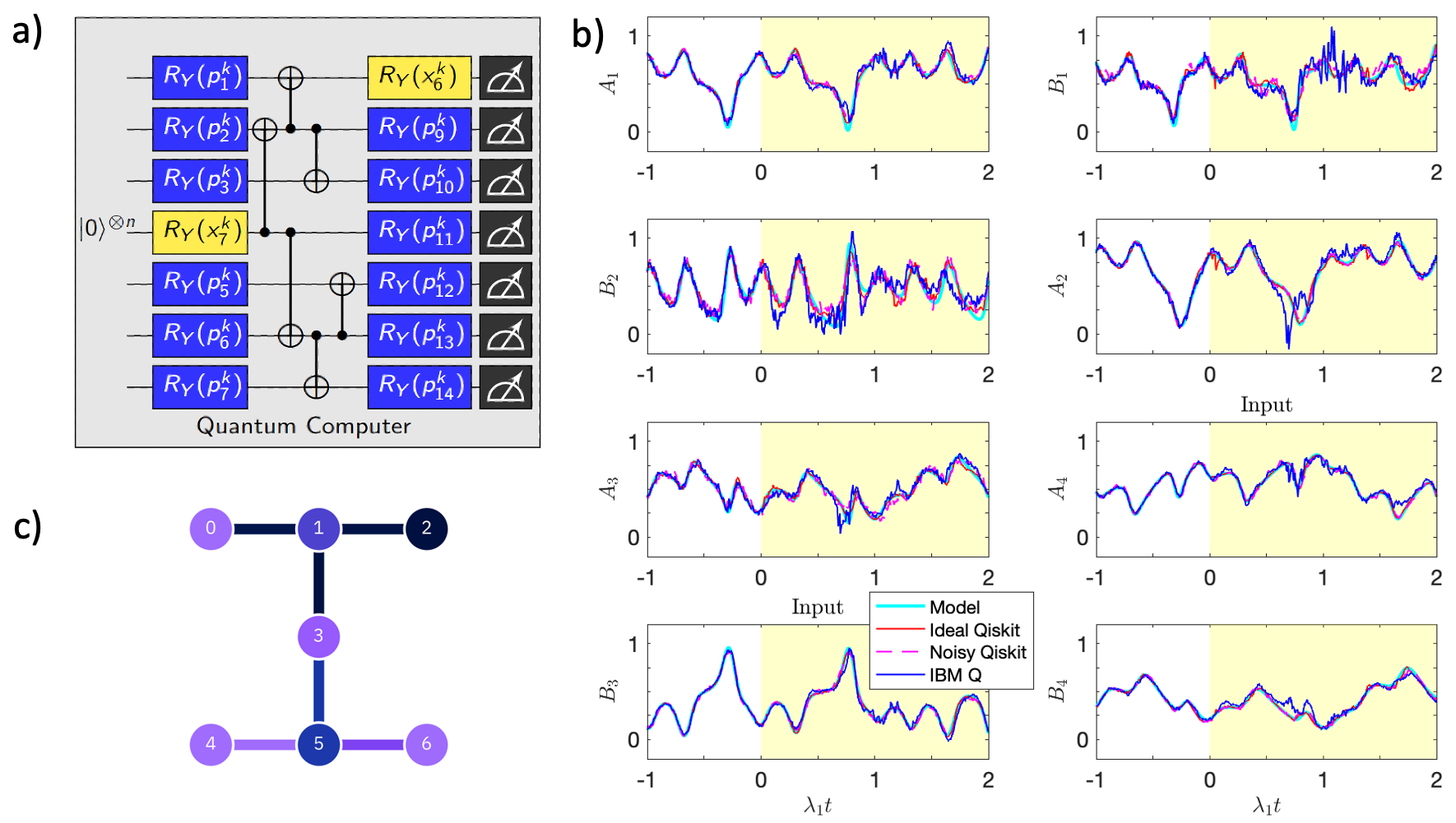}
\caption{Quantum reservoir computing model run of the 8-dimensional time-integrated Lorenz-type model an actual quantum device. (a) Sketch of the quantum reservoir which had to be reduced due to decoherence in comparison with the one that is displayed in Fig. \ref{fig1}. (b) Time series of the extended Lorenz model. We compare the ground truth (Model) with an ideal and noisy Qiskit simulator and the 7--qubit quantum computer (IBM Q). The number of training steps was again $N_{\rm train}=2000$ and the leaking rate $\varepsilon=0.2$. The two degrees of freedom that are continually available in the reconstruction phase are indicated. (c) Connection of the 7 qubits on the {\em ibm\_perth} quantum computer.}
\label{fig3}
\end{figure*}

\subsection{Implementation on an actual quantum device}

The 8-dimensional convection flow model is finally implemented on an actual noisy quantum device. Figure \ref{fig3}(a) displays the quantum reservoir for the implementation. The circuit depth on real devices is still rather limited by the decoherence of the elementary quantum gates that are installed in the form of microwave-controlled superconducting SQUIDs. The figure shows that we had to reduce the original three-block-structure of the quantum reservoir, which was used for the results that are shown in Figs. \ref{fig1} and \ref{fig2}, to one block. Two further steps were necessary: (1) Instead of one continually available variable in the reconstruction phase, we provide now $A_4(t)$ and $B_3(t)$. The reason is that the shallower circuit was found to be too noisy for the reconstruction of 7 degrees of freedom from one continually available degree of freedom. (2) Only 14 out of the 128 components of the reservoir state measurement vector ${\bm p}^t$ are fed back into reservoir together with the two degrees of freedom. This reduces the cost of loading data into a quantum register significantly. The total qubit number was limited to $n=7$. The studies were conducted on two devices, {\em ibmq\_ehningen}, a 27-qubit quantum computer in Germany, and {\em ibm\_perth}, a 7-qubit machine. Figure \ref{fig3}(c) displays the arrangement of the 7 qubits on {\em ibm\_perth} which was mainly used for the results in Fig. \ref{fig3}(b). The quantum computer {\em ibmq\_ehningen} was used for reservoir computing runs with $\varepsilon=1$, which gave a reduced network performance. The additional option to take the best callibrated 7 qubits out of 27 did not lead to significant improvements. Entanglement operations, e.g. by C-NOT gates, are only possible for qubits which are connected by the bars in Fig. \ref{fig3}(c). No error correction was performed.      

We backed up this investigation by two runs on the Qiskit simulator with the same configuration. One is the ideal simulator that has been used already before. The other simulation was done on a noisy Qiskit simulator for which you can prescribe the probabilities of measurement errors, here $p_m=0.05$, gate errors, here $p=0.1$, and qubit resets, here $p_r=0.03$. Values have been chosen such that they come close to those on real devices.  All environments are compared in Fig. \ref{fig3}(b). The data from the noisy Qiskit simulator and the real quantum device do partly deviate, but are found to follow the overall trend fairly well. This proves the concept of a hybrid QRCM for a classical dynamical system on a NISQ device. 
\begin{figure*}
\includegraphics[scale=0.55]{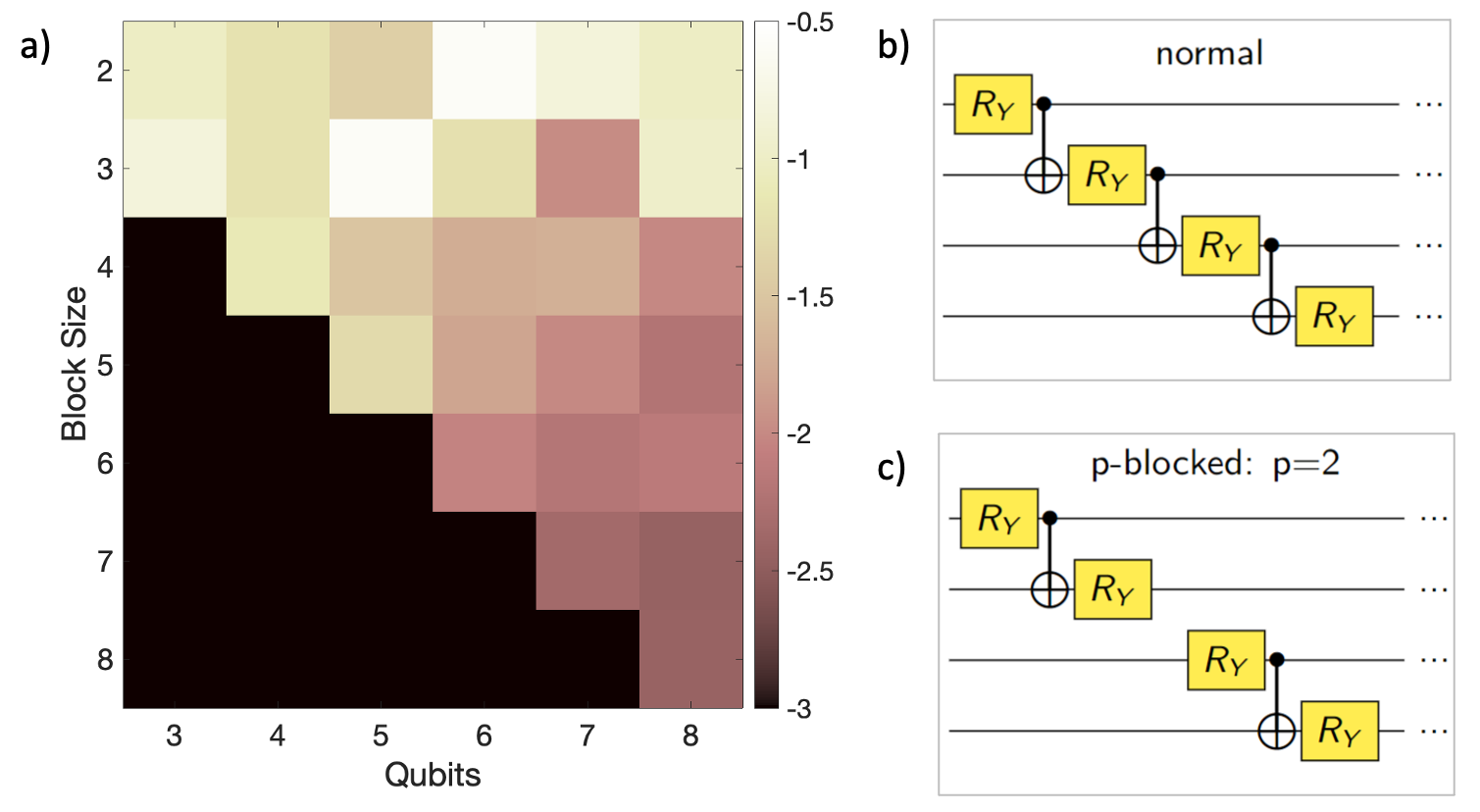}
\caption{Performance of the quantum reservoir computing model for different reservoir architectures. (a) Mean squared error on a logarithmic scale as a function of the total number of qubits and the size of the blocks of entangled qubits. The dark cells in the lower left stand for impossible decompositions. (b) Sketch of an example case. Fully entangled 4-qubit-quantum circuit which is the normal setting. (c) Two fully entangled 2-qubit blocks ($p=2$) build the 4-qubit-quantum circuit.}
\label{fig:blocked}
\end{figure*}
   
\subsection{Stepwise reduction of reservoir entanglement and quantum advantage}
Finally, we investigated if a simulation of the Lorenz-type model with the simpler quantum reservoir than from Fig. \ref{fig3}(a) is successful when the corresponding quantum circuit is decomposed into several $p$-qubit-blocks which are disentangled. If $p=n$ the circuit is fully entangled, for $p=1$ the $n$-qubit quantum state is fully separable; see appendix B for the definitions of both possible multi-qubit quantum states. The decomposition is illustrated in Figs. \ref{fig:blocked}(b,c). The rational behind this analysis, which we did with the ideal Qiskit simulator for $n=8$, is that a $p$-blocked structure might be simulated efficiently on a classical computer loosing the quantum advantage \cite{Josza2003}.   

In Fig. \ref{fig:blocked}(a), we summarize the MSE in a diagram for circuits with  $3\le n\le 8$ and possible block size $2\le p\le 8$. For example, $n=4$ and $p=3$ imply that a single qubit remains which is disentangled from the 3-qubit-block. In general, the number of blocks of size $p$ follows by $n_p=\lfloor n/p \rfloor$.  Each possible $p$-blocked reservoir configurations at $n$ qubits was trained and then run for 100 different realizations to gather statistics. Since the reduced structure of Fig. \ref{fig3}(a) is used, there is no $U({\bm \beta})$ block. To generate the 100 different realizations, we took 14 values $p_i$ randomly out of the 128 possible probabilities as input.  The block diagram shows that the MSE decreases when the block size is increased. We can conclude from this analysis that the entanglement of the qubits in the reservoir is essential for the performance of the QRCM. This is different compared to the classical reservoir which is a sparse network for which 20\% of the network nodes are actively only, but the number of perceptrons is by 2 to 3 order of magnitude larger in comparison to the number of qubits of the quantum reservoir. 
 
Finally, we estimate the number of operations for the CRCM and our QRCM. While most tasks have equal computational effort, our comparison is related to the second term in \eqref{MAIN_Eq} which provides the nonlinear activation. In the classical case, this requires mainly $\mathcal{O}((N_{\rm res}^c)^2)$ operations for the first term in the $\tanh(\cdot)$ activation; see ref. \cite{Heyder2021} for an analysis of the occupation of the reservoir matrix. Superscripts $c$ and $q$ of $N_{\rm res}$ stand for classical and quantum reservoir dimension, respectively. The second term with $N^c_{\rm res}\times N_{\rm in}$ operations is subdominant when $N_{\rm in}\ll N^c_{\rm res}$. In the quantum case, we have $\mathcal{O}(\xi n)$ gate operations for a shallow circuit as the one given in Fig. \ref{fig3}(a) that spans a reservoir of dimension $N_{\rm res}^q=2^n$. Here, the prefactor of $\xi\simeq 3$ is the approximate amount of operations for a single qubit. This has to be multiplied with the number of shots, which was set here to $K=2^{10+n}$. We compare the amount of operations for $N=N_{\rm res}^c=N_{\rm res}^q$ with $N$ being sufficiently large. Thus follows the inequality
\begin{equation}
    2^{10+n}\times \xi n < (2^n)^2 = 2^{2n} \,,
\end{equation}
or, in terms of states $N=2^n$,
\begin{equation}
    2^{10} \xi N \log_2(N) < N^2  \,,
\end{equation}
for having less reservoir operations of the quantum case. In order to show a quantum advantage for this framework rigorously, one needs to prove that the formula $K=2^{10+n}$ is still appropriate and that $\xi$ can still be choosen constant for increasing qubit number. Since the QRCM requires typically less nodes than the CRCM, i.e., $N_{\rm res}^q\ll N_{\rm res}^c$ as seen in Fig. \ref{fig_reg}, we expect that the QRCM might be able to outperform its classical counterpart, at least for the class of problems discussed here. It is clear that future investigations have to show if this is the case. 

\section{Summary and outlook}
The main objective of our present work was to show the feasibility of a hybrid quantum-classical reservoir computing model to predict and to reproduce the dynamical evolution of a classical, nonlinear thermal convection flow, on an actual quantum computer with up to 7 superconducting qubits. In a nutshell, quantum reservoir computing models are recurrent machine learning algorithms for which the reservoir state is built by a highly entangled tensor product quantum state that grows exponentially in dimension with the number of qubits. 
        
Our work showed that a quantum reservoir has a qubit number that is by about 2 orders of magnitude smaller than that of the perceptrons in a classical one. On the one hand, we could thus take advantage of the data compression capabilities of quantum machine learning algorithms where the dimension of the data space grows exponentially with the number of qubits which is essential for the modeling of higher-dimensional nonlinear dynamical systems. On the other hand, it shows that a classical reservoir state which is caused by a sparsely occupied network matrix of dimension $\gtrsim 10^3$ can be substituted by a highly entangled quantum state that is caused by the application of unitary transformations. The qubit number was $n<10$ in the present case. 

The study can be extended into several directions. It is clear that the present thermal convection flow model is still very low-dimensional and thus far away from convective turbulence. Our efforts should be considered as one first step to model real fluid flows on a quantum computer. It provides a possible route beside other directions, such as quantum embeddings of nonlinear dynamical systems by the Koopman operator framework \cite{Giannakis2022} or variational quantum algorithms for the direct solution of the equations of motion \cite{Gourianov2022}, see also ref. \cite{Bharadwaj2020} for further directions such as lattice Boltzmann methods. Extensions to higher-dimensional models are currently still limited by the technological capabilities of quantum computers. As the technological progress in this field is very fast, it can be expected that Galerkin models with significantly more modes will be modeled on upcoming devices with chips with a higher noise-resilience and lower error rates at the gates. The model that we applied here can be systematically extended towards turbulent convection, as discussed in detail in refs. \cite{Moon2017,Park2021}. A QRCM with $n\sim 10$ might thus be able to run a two-dimensional turbulent convection flow usable as a subgrid-scale superparametrization in a global circulation model \cite{Grabowski1999}.  

In the present work, we have not systematically optimized the circuit architectures for the different tasks. Further reductions of the number of gates caused always reduced prediction and reconstruction capabilities in the closed- and open-loop scenarios, respectively.  A possible route of research would thus be to compose the different quantum reservoirs more systematically from first principles, e.g. in the form of a multilayer tensorial network that potentially improves the performance of quantum algorithms on NISQ devices, see e.g. ref. \cite{Barratt2021}.   
    
\acknowledgments
This work is supported by the project no. P2018-02-001 "DeepTurb -- Deep Learning in and of Turbulence" of the Carl Zeiss Foundation and the Deutsche Forschungsgemeinschaft under grant no. DFG-SPP 1881. We acknowledge the use of IBM Quantum services for this work. The views expressed are those of the authors, and do not reflect the official policy or position of IBM or the IBM Quantum team. In this paper we used {\em ibmq\_ehningen} and {\em ibm\_perth}, which are IBM Quantum Falcon Processors. We thank Sachin Bharadwaj and Jean-Luc Thiffeault for helpful discussions. We acknowledge support for the publication costs by the Open Access Publication Fund of the Technische Universit\"at Ilmenau.

\appendix
\label{sec:materials_methods}

\section{Lorenz-type model of dimension 8}

In this appendix, we provide details of the derivation of reduced Lorenz-type models for thermally driven convection flows. The Lorenz-type models are obtained from eqns. \eqref{2d1} and \eqref{2d2} by means of the following finite expansions that satisfy the boundary conditions \eqref{bc}. They are given by
\begin{align}
\zeta(x,z,t)&=\sum_{i,j=1}^N c_\zeta A_{ij}(t) \Phi(\alpha_i x) \sin(\beta_j z)\,,\label{exp1}\\
\theta(x,z,t)&=\sum_{k,l=1}^M c_\theta B_{kl}(t) \Phi(\alpha_k x) \sin(\beta_l z)\,,\label{exp2}
\end{align}
with normalization prefactors $c_\psi , c_\theta$, the real amplitudes $\{A_{ij}(t),B_{kl}(t)\}$, and the wavenumbers 
\begin{align}
\alpha_k=k \alpha=\frac{2\pi k}{\Gamma}\quad\mbox{and}\quad
\beta_k=k\beta=k\pi\,,
\end{align}
with $k=0,1,2,\dots$ Here, $\Phi(x)=\{\cos(x), \sin(x)\}$. Inserting the expansions into the equations for $\zeta$ and $\theta$, sorting the terms with respect to the wavenumbers, and truncating higher wavenumbers leads to closed systems of coupled nonlinear ordinary differential equations. In the present work, we consider Lorenz-type models up to order 8 ($N=M=4$); higher-dimensional models have been investigated for example in refs. \cite{Moon2017, Gluhovsky2002}. In detail, we take the expansions 
\begin{align*}
\zeta(x,z,t)&=c_\zeta [A_1(t) \sin(\alpha x) \sin(\beta z)+
                     A_2(t) \sin(\beta z) \nonumber \\
            &+A_3(t) \cos(\alpha x) \sin(2\beta z) +A_4(t)\sin(3\beta z)]\,,\\
\theta(x,z,t)&=c_\theta [\sqrt{2} B_1(t) \cos(\alpha x) \sin(\beta z)+
                   B_2(t) \sin(2\beta z) \nonumber \\
            &+B_3(t) \sin(\alpha x) \sin(2\beta z) + B_4(t) \sin(4\beta z)]\,.
\end{align*}
The normalization constants are 
\begin{align}
c_\zeta&=\sqrt{2}\,\frac{\alpha^2+\beta^2}{\alpha\beta}\,,\\ c_\theta&=\frac{(\alpha^2+\beta^2)^3}{\alpha^2\beta{\rm Ra}}\,.
\end{align} 
The resulting system of nonlinear coupled ordinary differential equations is given by 
\begin{align}
\frac{dA_1}{d\tau}&=\sigma \left(B_1 - A_1\right) - \frac{3\beta^2 + \alpha^2}{\sqrt{2}(\alpha^2 + \beta^2)} A_2A_3 \nonumber \\ &+ \frac{3\alpha^2-15\beta^2}{\sqrt{2}(\alpha^2 + \beta^2)} A_3A_4\,,\\
\frac{dA_2}{d\tau}&=-\frac{\sigma b}{4} A_2 -\frac{3}{2\sqrt{2}} A_1A_3\,,\\
\frac{dA_3}{d\tau}&=-\sigma \frac{\alpha^2 + 4\beta^2}{\alpha^2 + \beta^2} A_3 - \sigma \frac{\alpha^2 + \beta^2}{\sqrt{2}(4\beta^2 + \alpha^2)}B_3 \nonumber \\ &+ \frac{\alpha^2}{\sqrt{2}(\alpha^2 + \beta^2)} A_1A_2 + \frac{24\beta^2 -3\alpha^2}{\sqrt{2}(4\beta^2 + \alpha^2)} A_1A_4\,,\\
\frac{dA_4}{d\tau}&=-\frac{9 \sigma b}{4} A_4-\frac{1}{2\sqrt{2}} A_1A_3\,,\\
\frac{dB_1}{d\tau}&=-B_1 + rA_1 + A_1B_2 + \frac{1}{2}A_2B_3 +  \frac{3}{2}A_4B_3\,,\\
\frac{dB_2}{d\tau}&= - bB_2 - A_1 B_1 \,,\\
\frac{dB_3}{d\tau}&= -\frac{\alpha^2 + 4\beta^2}{\alpha^2 + \beta^2}B_3 -A_2B_3 + \sqrt{2}r A_3 + 3A_4B_1 \nonumber\\ &- 2\sqrt{2} A_3B_4\,,\\
\frac{dB_4}{d\tau}&=-4b B_4 + \frac{3\sqrt{2}}{4} A_3B_3\,,
\end{align}
with $b=4\beta^2/(\alpha^2+\beta^2)$, ${\rm Ra}_c=(\alpha^2+\beta^2)^3/\alpha^2$, and the rescaled time $\tau=(\alpha^2+\beta^2)t$. The Lorenz 63 model \cite{Lorenz1963} is recaptured for $A_2=A_3=A_4=0$ and $B_3=B_4=0$.

A primary linear instability of the convection flow, which initiates fluid motion due to a sufficiently large temperature difference between the bottom and the top, takes place at a critical Rayleigh number ${\rm Ra}_c=27\pi^4/4$ when free-slip boundary conditions hold at the top and bottom \cite{Chandrasekhar1961}. The 1st parameter $\sigma$ is the Prandtl number. The 2nd parameter $r$ in the Lorenz-type models is then defined as $r={\rm Ra}/{\rm Ra}_c>1$. The 3rd parameter $b$ is connected to the aspect ratio $\Gamma$ of the fluid volume that is considered. In detail,
\begin{equation}
b=\frac{4\Gamma^2}{4 + \Gamma^2}\,.
\end{equation}
If $b=8/3$ then the aspect ratio is $\Gamma=\mbox{length/height}=2\sqrt{2}$ which corresponds to the critical wavelength of plane wave perturbations of the quiescent equilibrium of the convection flow. In other words, at this wavelength the thermal convection flow becomes linearly unstable first.

\section{Qubits and tensor product spaces}
In this appendix, we briefly summarize some basic definitions of quantum computing. For more details we refer to the textbook of Nielsen and Chuang \cite{Nielsen2010} or a review by Bharadwaj and Sreenivasan \cite{Bharadwaj2020}. While a single classical bit can take two discrete values, namely $\{0,1\}$ only, a single quantum bit (in short qubit) is a superposition of two basis states in the vector space $\mathbb{C}^2$ which can take any state on the surface of a (Bloch) sphere
\begin{equation}
|q_1\rangle=c_1|0\rangle+c_2|1\rangle=c_1
\left(
\begin{array}{c}
1 \\ 0\\
\end{array}\right)
+c_2
\left(
\begin{array}{c}
0 \\ 1\\
\end{array}\right)\,, \label{Eq:B1}
\end{equation}
with $c_1, c_2\in \mathbb{C}$ and $\sqrt{|c_1|^2+|c_2|^2}=1$ . Vectors $|0\rangle$ and $|1\rangle$ are the basis vectors in Dirac notation \cite{Nielsen2010}. A two-qubit state vector is the tensor product of two single-qubit vectors, 
\begin{equation}
|q_2\rangle=|q_1\rangle\otimes |q_1^{\prime}\rangle\,.
\end{equation}
The basis of this tensor product space is given by 4 vectors: $|j_1\rangle=|0\rangle\otimes |0\rangle$, $|j_2\rangle=|0\rangle\otimes |1\rangle$, $|j_3\rangle=|1\rangle\otimes |0\rangle$, and $|j_4\rangle=|1\rangle\otimes |1\rangle$. An $n$-qubit quantum state, which is given by
\begin{equation}
   \ket{\Psi} = \sum\limits_{k=1}^{2^n} c_k \ket{j_k} \hspace*{1em} \text{with} \hspace*{1em} \sum\limits_{k=1}^{2^n} |c_k|^2 = 1\,,
\end{equation}
is called {\em fully separable} if it can be written as
\begin{equation}
   \ket{\Psi} = \overset{n}{\underset{i=1}{\bigotimes}} \,\ket{q_i} \,,
\end{equation}
where $\ket{q_i}$ are single qubit quantum states given by eq. (B1). It is called {\em separable} if a tensor product decomposition of $\ket{\Psi}$ into blocks is possible with at least one multi-qubit quantum state $\ket{q_i}$, that is not fully separable. Not separable multi-qubit quantum states are called {\em entangled}. An $n$-qubit quantum state is called {\em fully entangled} if no subspace of separable qubits exists.

\section{Classical reservoir computing model}
In this appendix, we provide details about the reservoir computing approach, a recurrent supervised machine learning algorithm, which here is implemented in the form of an echo state network with a leaking rate $\varepsilon$. The {\em training} of the RCM proceeds as follows. The dynamical system state at time $t$, which is denoted more compactly as 
\begin{equation}
    {\bm x}^t=\{A_1,...A_N,B_1,...,B_M\}\in \mathbb{R}^{N_{\rm in}}\,,
\end{equation}
is mapped to the reservoir state ${\bm \psi}^t$ via the randomly initialized input weight matrix $W^{\rm in}\in \mathbb{R}^{N_{\rm res}\times N_{\rm in}}$. Here, $N_{\rm res}\gg N_{\rm in}$ is the reservoir dimension. The reservoir state ${\bm \psi}^t$ is updated as follows \cite{Jaeger2004,Pathak2018,Pandey2020}, see eq. \eqref{rcm_train0} in the main text,
\begin{equation}
	\label{rcm_train}
	{\bm \psi}^{t+1}= (1-\varepsilon){\bm \psi}^t + \varepsilon\tanh\left[W^{\rm in}{\bm x}^t + W^{\rm r} {\bm \psi}^{t}\right].
\end{equation}
This update rule comprises external forcing by the inputs ${\bm x}^t$ as well as a self-interaction with the reservoir state ${\bm \psi}^{t}$. The two terms on the right hand side of \eqref{rcm_train} are combined by the leaking rate $\varepsilon$. The hyperbolic tangent $\tanh\left(\cdot\right)$ is the nonlinear activation function of each reservoir node. The randomly initialized matrix $W^{\rm r}$ represents the reservoir, a sparse random network of neurons \cite{Jaeger2002}. Thus the leaking rate $\varepsilon\in \left(0,1\right]$ moderates the linear and nonlinear contributions. The updated reservoir state ${\bm \psi}^{t+1}$ is mapped via the output matrix $W^{\rm out}\in \mathbb{R}^{N_{\rm in}\times N_{\rm res}}$ to form the reservoir output ${\bm x}^{t+1}\in \mathbb{R}^{\rm N_{\rm in}}$
\begin{equation}
	\label{eq:rcm_out}
	{\bm x}^{t+1}=W^{ \rm out}{\bm \psi}^{t+1}\,.
\end{equation}
The elements of $W^{\rm out}$ have to be computed. Therefore, a set of $T$ training data instances $\{{\bm x}^{t+1},{\bm x}_{\rm tg}^{t+1}\}$, where $t= -T, -T+1, ..., -1$, needs to be prepared. The target output ${\bm x}_{\rm tg}^{t+1}$ (also denoted as ground truth (GT)) represents the desired output that the RCM should produce for the given input ${\bm x}^t$. The resulting data pairs are assembled into a mean squared cost function $C\left(W^{\rm out}\right)$ with a Tikhonov regularization term which is given by  
\begin{equation*}
	C(W^{\rm out}) =\frac{1}{T}\sum_{t = -T}^{-1} |{\bm x}^t-{\bm x}_{\rm tg}^{t}|^2
	+\gamma {\rm Tr} \left(W^{\rm out}W^{\rm out^T}\right)\,, 
	\label{eq:rcm_train1}
\end{equation*}
and has to be minimized corresponding to $W^{{\rm out}\ast} = \arg\min C(W^{\rm out})$. Superscript T denotes the transposed. The regularization parameter $\gamma>0$ avoids overfitting \cite{Goodfellow2016}. The optimized output matrix is given by
\begin{eqnarray}
\label{eq:rcm_fitted_wout}
W^{{\rm out}\ast} = U_{\rm tg}R^{\rm T}\left(RR^{\rm T}+\beta \mathbb{I}\right)^{-1}
\end{eqnarray}
where $\mathbb{I}$ is the identity matrix. $U_{\rm tg}$ and $R$ are matrices where the $t$-th column is the target output ${\bm x}_{\rm tg}^{t}$ and reservoir state ${\bm \psi}^t$, respectively. The optimization of the output weights thus becomes computationally inexpensive. The hyperparameters of the classical RCM are $N_{\rm res}$, $\varepsilon$, $\gamma$, the reservoir density $D$, and the spectral radius $\rho(W^{\rm r})$. 

Once the output weights are optimized and the hyperparameters are tuned the RCM can run either in the {\em prediction} (closed-loop scenario) or {\em reconstruction mode} (open-loop scenario). Equation \eqref{rcm_train} changes in the closed-loop scenario to 
\begin{align}
	\label{eq:rcm_activation}
	\mathbf{\bm \psi}^{t+1}= &(1-\varepsilon){\bm \psi}^{t}+\varepsilon\tanh\left[W^{ \rm in}W^{{\rm out}\ast}{\bm \psi}^t + W^{\rm r}{\bm \psi}^{t}\right].
\end{align}
Now the RCM can work independently of training input. The prediction for the dynamical system follows by ${\bm x}^{t+1}=W^{{\rm out}\ast}{\bm\psi}^{t+1}$. Equation \eqref{rcm_train} remains the same in the open-loop scenario, except that the continually available input vector is very low-dimensional in this regime, see Fig. \ref{fig0}. The full state reconstruction follows again by ${\bm x}^{t+1}=W^{{\rm out}\ast}{\bm\psi}^{t+1}$. The latter case is also called one-step prediction since ${\bm x}^{t+1}$ is not used as a new input for ${\bm x}^{t+2}$, differently to the former prediction mode.

\section{NARMA-2 model and Mackey-Glass equation for different leaking rates}
 In this appendix, we demonstrate the necessity of $\varepsilon<1$ for two common reservoir computing benchmark cases. The first case is the NARMA model, an input-output model class with input $u_k$ with $k\in\mathbb{N}$ given by
\begin{equation}
u_k = 0.1 [ \sin(2\pi\alpha k)\sin(2\pi\beta k)\sin(2\pi\gamma k)+1 ]\,.
\end{equation}
Here, $\alpha=2.11/T$, $\beta=3.73/T$, $\gamma=4.11/T$, and $T=100$. The output $y_k$ is then given by the following iteration rule
\begin{equation}
y_{k+1} = 0.4y_k+0.4y_ky_{k-1}+0.6 u_k^3+0.1\,.
\end{equation}
We use $y_0=y_1=0.19$. The recursive character of the discrete time series can be enhanced by adding further terms from the past. Here, we take a NARMA-2 model since $y_{k+1}$ depends on $y_k$ and $y_{k-1}$.

\begin{figure}[h]
    \centering
    \includegraphics[scale=0.23]{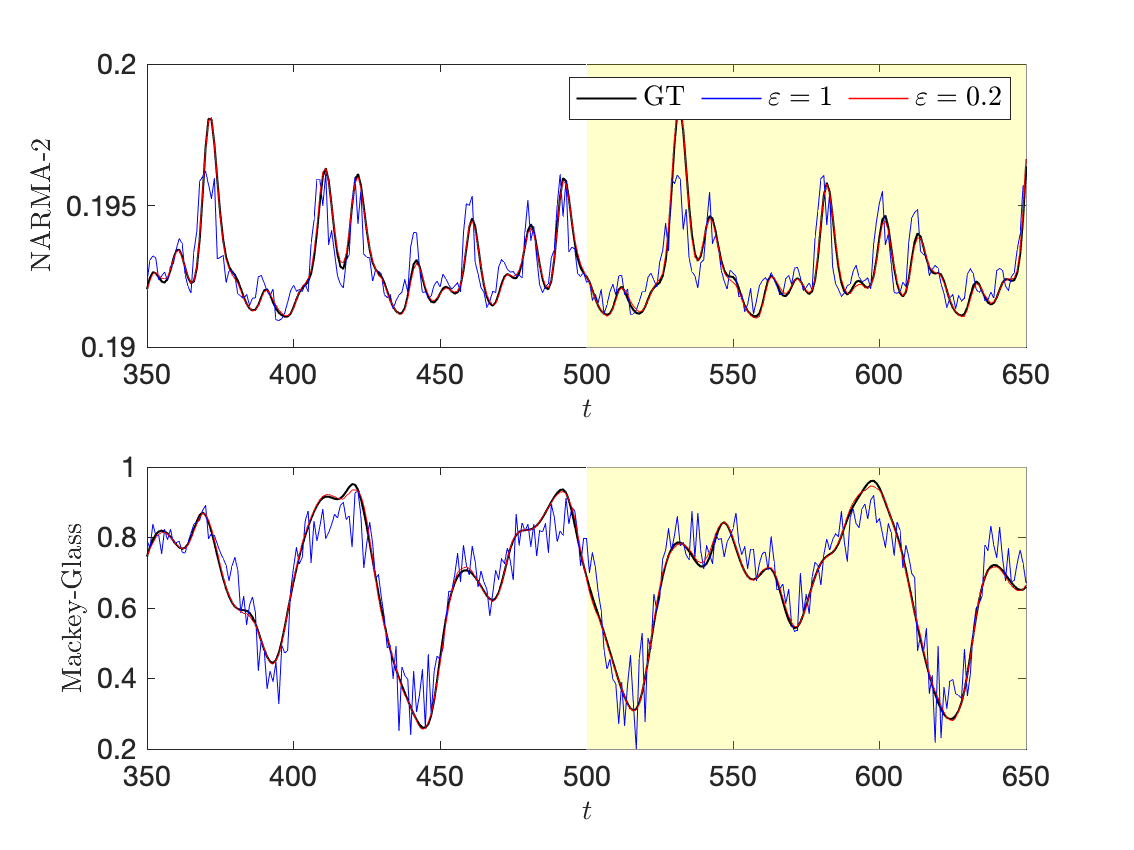}
    \caption{Open-loop prediction with the quantum circuit structure of Fig. \ref{fig1}(b) for the time-discrete NARMA-2 model with 4 qubits (top) and the time-continuous Mackey-Glass equation for 5 qubits (bottom). GT is the ground truth. Leaking rates in the legend hold for both panels. Training ends at time step $k=500$. The open-loop one-step prediction is marked by the yellow shaded background. }
    \label{fig:NARMA}
\end{figure}

The second case is the Mackey-Glass equation, a nonlinear time-delay differential equation, which is given by 
\begin{equation}
\frac{dx(\tau)}{d\tau} = 
\frac{\beta \alpha^n x(\tau-T) }{\alpha^n+x(\tau-T)^n} - \gamma x(\tau)\,.
\end{equation}
Here, $\alpha=1$, $\beta=2$, $\gamma=1$, and $T=2$. The time $\tau$ is measured here in multiples of the time step width $\Delta\tau=0.1$, i.e., $\tau=k\Delta\tau$ with $k\in \mathbb{N}$. Figure \ref{fig:NARMA} compares the aforementioned benchmarks for two leaking rates which were processed with our QRCM, either with $\varepsilon=1$ or $\varepsilon=0.2$. All runs were done in Qiskit. The first 100 time steps are used for washout, the subsequent 400 time steps for training. We clearly observe a significant improvement of the performance of the hybrid quantum-classical reservoir computing model with a leaking rate of $\varepsilon<1$. The QRCM prediction with $\varepsilon=0.2$ follows the ground truth nearly perfectly.

\bibliographystyle{unsrt}
\bibliography{references}

\end{document}